\documentclass[twocolumn,showpacs,prd,aps,nofootinbib,showkeys,unsortedaddress,10pt]{revtex4-2}

\usepackage{amsmath}
\usepackage{mathtools}
\usepackage{amssymb}
\usepackage{slashed}
\usepackage{hyperref} 
\usepackage{color}
\usepackage{comment}
\usepackage{simpler-wick}

\usepackage{tikz-feynhand}

\newcommand{\be}{\begin{equation}}
\newcommand{\ee}{\end{equation}}
\newcommand{\eq}[1]{Eq.\ (\ref{#1})}

\newcommand{\appref}[1]{App.\ \ref{#1}}

\newcommand{\infinity}{\infty}

\newcommand{\ess}{\hspace{0.1em}}

\newcommand{\msbar}{$\overline{\mathrm{MS}}$}

\newcommand{\im}{\mathrm{Im}\,}
\renewcommand{\Re}{\mathrm{Re}\,}
\newcommand{\re}{\mathrm{Re}\,}

\definecolor{darkgreen}{rgb}{0,.7,0}
\definecolor{linkblue}{rgb}{0.,0.,0.9333}

\DeclareMathOperator{\sinc}{sinc}

\hypersetup{
    pdffitwindow=true,      
    pdfauthor={W. A. Horowitz},     
    pdfnewwindow=true,      
    bookmarksnumbered=true,
    colorlinks=true,       
    linkcolor=red,          
    citecolor=darkgreen,        
    filecolor=magenta,      
    urlcolor=cyan,           
    menucolor=blue,
    baseurl=http://www.arxiv.org/abs/,
}

\AtBeginDocument{\setlength\abovedisplayskip{5pt}}
\AtBeginDocument{\setlength\belowdisplayskip{5pt}}

\begin{document}

\title{Finite System Size Correction to NLO Scattering in \texorpdfstring{$\phi^4$}{phi4} Theory}

\author{W.\ A.\ Horowitz}%
\email{wa.horowitz@uct.ac.za}
\affiliation{%
 Department of Physics, University of Cape Town, Rondebosch 7701, South Africa
}%

\author{J.\ F.\ Du Plessis}%
\email{23787295@sun.ac.za}
\affiliation{%
 Department of Physics, University of Stellenbosch, Matieland, 7602, South Africa
}%

\date{\today}

\begin{abstract}
    We compute $2\rightarrow2$ scattering in massive $\phi^4$ theory on $\mathbb R^{1,m}\times T^n$ to NLO.  
    We perform the calculations using ``denominator regularization'' instead of the usual dimensional regularization, which allows for asymmetric configurations of the $T^n$.  We give a transparent derivation of and equation for the analytic continuation of the generalized Epstein zeta function.  We show that the Optical Theorem is satisfied and generalize a conjecture by Hardy on square counting functions.      We comment on the implications.
\end{abstract}
\maketitle

\section{Introduction}
Measurements from RHIC and LHC show that signs of quark-gluon plasma (QGP) formation seen in large nucleus-nucleus collisions are also present in high multiplicity p+p and p+A collisions \cite{Abelev:2012ola,Aad:2012gla,Adare:2013piz,Aad:2015gqa}.  The distribution and correlations between low momentum particles in these small system collisions are well described using relativistic, nearly-inviscid hydrodynamics using an equation of state computed using lattice QCD and extrapolated to infinite spatial volume \cite{Bzdak:2013zma,Weller:2017tsr}.  The interpretation of the latter is that the medium produced in these high multiplicity small systems is a nearly invsicid QGP of the same nature as that produced in large system A+A collisions.

A recent investigation of the finite size effects in a massless free scalar thermal field theory with Dirichlet boundary conditions showed that finite size effects can effectively mimic the effects of temperature dependence on the phase structure of full QCD \cite{Mogliacci:2018oea}.  40\%+ corrections to the usual thermodynamic quantities of pressure, entropy, etc.\ were found for systems of the size of p+p collisions; even for systems of the size of mid-central A+A collisions showed $\sim10\%$ corrections.  Quenched lattice QCD calculations with periodic boundary conditions confirmed the importance of finite size effects in systems of asymmetric size \cite{Kitazawa:2019otp}.

The equation of state---equivalently the speed of sound or the trace anomaly---plays a critical role in hydrodynamics simulations of high multiplicity relativistic hadronic collisions.  Surprisingly, despite the breaking of conformal symmetry due to the presence of Dirichlet boundary conditions, we find that the free massless scalar field theory yields a trace of the energy momentum tensor that is identically 0 \cite{Horowitz:2021dmr}.  Presumably, then, any trace anomaly in a massless theory in a finite-sized system must come from running coupling effects.  The effect of the finite size correction to the trace anomaly in QCD was very roughly estimated by using the finite size correction to the coupling as calculated in a massive scalar theory \cite{Montvay:1994cy}.  It was seen that the finite size corrections dramatically reduced the size of the trace anomaly \cite{Horowitz:2021dmr}.  Such a large reduction in the trace anomaly would have a significant impact on the extracted sheer and bulk viscosities from comparing hydrodynamics simulations to data.

We're therefore interested in computing analytically the effect of a finite system size on the trace anomaly of QCD induced through the finite size effect on the QCD coupling.  This is a significant challenge that will require understanding several important techniques.  The two most important challenges will be to understand how to regularize and renormalize the thermal field theory in a finite size system and to include the effect of torons, non-trivial vacuum gauge configurations on a torus \cite{tHooft:1979rtg,Coste:1985mn}.  

This work provides a step in the direction of the first challenge by computing the finite size correction to the running coupling in massive $\phi^4$ theory for $2\rightarrow2$ scattering.  In order to perform this computation, we introduce a technique that we will call ``denominator regularization.''  While one can formulate dimensional regularization on a hypercube of equal sides \cite{Zinn-Justin:2002ecy}, denominator regularization is a more natural procedure and also allows for asymmetric spaces.  The freedom to have asymmetric spaces allows us to smoothly capture results for, e.g., $n=1,\,2,$ and 3 compact dimensions when starting from an $\mathbb R\times T^3$ space.  (More broadly, denominator regularization provides an alternative to the heat kernel or zeta function regularization in curved spacetimes \cite{Hawking:1976ja} and avoids the complications with dimensional regularization and the representations of the Lorentz group for field theories of spin greater than 0.)  Following \cite{Elizalde:1997jv} we then derive an analytic continuation of the generalized Epstein zeta function that is of critical value when employing denominator regularization and is exceptionally well suited to future thermal field theory studies in small systems.  We then apply this analytic continuation to the problem at hand.  We perform a non-trivial check by confirming that our $2\rightarrow2$ amplitude at NLO satisfies the Optical Theorem.  This check suggests to us a generalization of a conjecture of Hardy on the square counting function \cite{Hardy1978RamanujanTL}.  

\section{Finite Size Corrections}
Consider a real scalar field theory with $n$ directions periodically identified and $m$ directions of infinite extent.  Let the $i^\mathrm{th}$ compact spatial dimensions have size $[-\pi L_i,\pi L_i]$, where the $L_i$ for different $i$ do not have to be equal. If we restrict ourselves to three periodic spatial directions and no spatial directions of infinite extent, $n=3$ and $m=0$, we may immediately write down the quantity needed to evaluate the NLO correction to $2\rightarrow2$ scattering.  We will see that this setup also captures the $n<3$ physics.  Defining $p\equiv p_A+p_B$, where $p_A$ and $p_B$ are the incoming momenta, and $V(p^2)$ by
\newcommand{\hi}{0.5}
\newcommand{\wi}{0.25}
\newcommand{\wid}{0.55}
$
    (-i\lambda)^2iV(p^2)
    \equiv
    \begin{tikzpicture}[baseline=(c.base)]
        \begin{feynhand}
            \setlength{\feynhandarrowsize}{4pt}
            \setlength{\feynhanddotsize}{0mm}
            \vertex (a) at (-\wid,\hi) {};
            \vertex (b) at (-\wid,-\hi) {};
            \vertex [dot] (c) at (-\wi,0) {};
            \vertex [dot] (d) at (\wi,0) {};
            \vertex (e) at (\wid,\hi) {};
            \vertex (f) at (\wid,-\hi) {};
            \draw (a) to (c);
            \draw (b) to (c);
            \draw (c) to [out = 65, in = 115, looseness = 1.75] (d);
            \draw (c) to [out = 295, in = 245, looseness = 1.75] (d);
            \draw (d) to (e);
            \draw (d) to (f);
        \end{feynhand}
    \end{tikzpicture}$ 
we have after combining denominators and Wick rotating
\begin{align}
    V(p^2,\{L_i\})
    & = -\frac12\int_0^1dx\int\frac{d\ell_E^0}{2\pi}\sum_{\vec k\in\mathbb Z^3} \nonumber\\
    \label{eq:one}
    & \qquad\qquad \frac{1}{(2\pi)^3L_1L_2L_3}\frac{1}{[\ell_E^2+\Delta^2]^{2}}, 
\end{align}
where $\Delta^2 \equiv -x(1-x)p^2 + m^2 - i\varepsilon$ and $\ell^\mu_E = (\ell_E^0,\frac{k^i}{L_i}+x\ess p^i)^\mu$.  The above is UV divergent.  To capture the divergence we introduce \emph{denominator regularization}.  Instead of analytically continuing the number of spacetime dimensions, we allow the \emph{power} of the denominator in the loop integral to be a variable and analytically continue to the log divergent value of 2.  To keep $V$ dimensionless we must simultaneously introduce a dimensionful scale $\mu$.  We thus are interested in
\begin{multline}
    V(p^2,\{L_i\};\mu,\epsilon)
    = -\frac12\int_0^1dx\int\frac{d\ell_E^0}{2\pi}\sum_{\vec k\in\mathbb Z^3} \\ 
    \frac{1}{(2\pi)^3L_1L_2L_3}\frac{\mu^{2\epsilon}}{[\ell_E^2+\Delta^2]^{2+\epsilon}}.
\end{multline}
Notice how one cannot as in the infinite size case simply shift the spatial integration to remove the $+x\ess p^i$ shift in $\ell_E^\mu$.  Evaluation of the $\ell_E^0$ integral yields
\begin{multline}
    V(p^2,\{L_i\};\mu,\epsilon)
    = -\frac12\frac{1}{2\pi}\frac{1}{(2\pi)^3L_1L_2L_3}\int_0^1dx \\ \times\frac{\sqrt\pi\Gamma\big(\frac32+\epsilon\big)}{\Gamma(2+\epsilon)}
    \sum_{\vec k\in\mathbb Z^3}\frac{\mu^{2\epsilon}}{\left( \sum_{i=1}^3\big(\frac{k^i}{L_i}+x\ess p^i\big)^2 + \Delta^2 \right)^{\frac32+\epsilon}}. \nonumber
\end{multline}

Our result includes a generalized Epstein zeta function \cite{Kirsten:1994yp},
\begin{align}
    \label{e:epsteinzeta}
    \zeta(\{a_i\},\{b_i\},c;s) & \equiv \sum_{\vec n \in \mathbb Z^p} \big[ a_i^2 n_i^2 + b_i n_i + c \big]^{-s},
\end{align}
where repeated indices are assumed summed over.  The generalized Epstein zeta function converges for $s>d$.  As per usual we wish to isolate the pole occurring at $s=d$ and determine the finite remainder.  To do so, we utilize the Poisson summation formula to provide an analytic continuation of the generalized Epstein zeta function; we detail the derivation in \appref{s:appsum}.  We may immediately apply \eq{e:epsteinanalytic} with $s=\frac 32+\epsilon$ to find
\begin{widetext}
    \begin{align}
        V(p^2,\{L_i\};\mu,\epsilon) = -\frac12\frac{1}{(4\pi)^2}\int_0^1dx\left\{ \frac1\epsilon - 1 + \ln\frac{\mu^2}{\Delta^2} 
        +2\sideset{}{'}\sum_{\vec m\in\mathbb Z^3}e^{-2\pi\ess i \ess x \sum m_i p^i L_i} K_0\Big( 2\pi\sqrt{\Delta^2\sum m_i^2L_i^2} \Big)  \right\} + \mathcal O(\epsilon),
    \end{align}
\end{widetext}
where the suppressed limits of the sums run from $i=1\ldots3$.  It's interesting that the $-1$ of the finite part from denominator regularization is identical to the $-1$ that one finds when regularizing through an explicit UV cutoff.  One may find similar expressions using \eq{e:epsteinanalytic} for different numbers of spatial dimensions; for $n<3$ there's no divergence.  

We may modify the usual \msbar{} convention to have the counterterm absorb the ubiquitous $-1$ from denominator regularization.  Then the renormalized NLO contribution to $2\rightarrow2$ scattering in 3 periodic spatial dimensions is
\begin{multline}
    \label{e:renormfiniteV}
    \overline V(p^2,\{L_i\};\mu) = -\frac12\frac{1}{(4\pi)^2}\int_0^1dx\left\{ \ln\frac{\mu^2}{\Delta^2} \right. \\
        \left. +2\sideset{}{'}\sum_{\vec m\in\mathbb Z^3}e^{-2\pi\ess i \ess x \sum m_i p^i L_i} K_0\Big( 2\pi\sqrt{\Delta^2\sum m_i^2L_i^2} \Big)  \right\},
\end{multline}
and the counterterm is unchanged from the $\mathbb R^{1,3}$ case.  Notice that in the denominator regularization case we have completely removed the $\epsilon$ dependence in the renormalized $\overline V$.

Since asymptotically $K_0(z)\sim \exp(-z)/\surd z$ we see that the finite size corrections naturally go to zero as the system size grows and the result converges to the $\mathbb R^{1,3}$ limit $\sim\ln\mu^2/\Delta^2$.  We see that we also obtain the results for $n=0,\,1,\,$ and $2$ since we may take the associated $L_i\rightarrow\infinity$ limit smoothly.  Notice further that the UV divergence is unaffected by the finite size corrections.  We should have expected this lack of sensitivity of the UV divergence to the finite system size, since a finite system size acts as an IR cutoff; the infinitely small distances probed at the infinite UV are insensitive to the global existence of a finite-sized edges or periodic boundary conditions for the manifold (effectively) infinitely far away.  As a result, a leading logarithmic analysis such as from an application of the Callan-Symanzik equation won't be able to capture the finite size effects on the running coupling; rather, we must explicitly perform the resummation of the 1PI diagrams to see the subleading $1/L$ corrections to the running coupling.  Even though this analysis is subleading log in the limit of large $p$, we're interested in the momentum region in which the finite size effects aren't vanishingly small; i.e., we're interested in the case of $p\lesssim1/L$.

\section{Unitarity Check}
For self-consistency we should find that $2\ess\im\mathcal M = \sigma_{tot}$.  In general, to NLO, 
\begin{multline}
    2\ess\im\mathcal M
     = -2\lambda^2\ess\im\big( \overline V(s,\{L_i\};\mu) \\ + \overline V(t,\{L_i\};\mu) + \overline V(u,\{L_i\};\mu) \big), 
\end{multline}
where $\overline V(p^2,\{L_i\};\mu)$ is given by \eq{e:renormfiniteV}.  As noted in \appref{s:appsum}, one may organize the sum for the finite size correction such that the phases are only cosines.  Therefore the only contribution to the imaginary part of the amplitude may come from values of $x$ such that $\Delta^2<0$, in which case there are contributions from evaluating the logarithm of negative numbers and from evaluating the modified Bessel function for arguments with an imaginary part.  Since $t$ and $u$ are non-positive, we must therefore have that $\im\mathcal M$ only comes from $\overline V(s,\{L_i\};\mu)$.  We will in general work in the center of mass frame, in which case $p^i = 0$ for the $s$ channel $\overline V(s,\{L_i\};\mu)$.  Recall from the $\mathbb R^{1,3}$ case that $\re\Delta^2<0$ for $s>4m^2$ and $x_-<x<x_+$, where $0<x_\pm<1$ are given by $x_\pm = \frac12\left[ 1\pm\sqrt{1-\frac{4m^2}{p^2}}\right]$.  We self-consistently align the branch cuts of $\arg$, $\log$, and $K_0$ along the negative real axis.  Then the small imaginary part from the propagators in the loop means that for $s>4m^2$ and $x_-<x<x_+$ we have that $\Delta^2$ is in the third quadrant of the complex plane; thus $\sqrt{\Delta^2}$ is in the fourth quadrant of the complex plane.  Therefore for $x_-<x<x_+$ we have, for $\Delta^2\equiv-x(1-x)s+m^2-i\varepsilon$ 
\begin{align}
    \im K_0\big( 2\pi\sqrt{\Delta^2\sum m_i^2L_i^2} \big) 
    = \frac\pi2J_0\big(2\pi|\Delta|\sqrt{\sum m_i^2 L_i^2}\big), \nonumber
\end{align}
where $J_0$ is the usual Bessel function of the first kind, and we've dropped the irrelevant terms linear and higher order in $\varepsilon$ on the right hand side (and can take $|\Delta| = x(1-x)s - m^2$).  

Defining $\tilde Q \equiv \sqrt{1-(4m^2/s)}$ we have
\begin{align}
    2\, \im\mathcal M 
    & = \frac{\lambda^2}{16\pi}\tilde Q\theta(\tilde Q^2) \bigg[ 1 \nonumber\\ 
    & \qquad + \frac{1}{\tilde Q}\sideset{}{'}\sum_{\vec m\in\mathbb Z^n} \int_{x_-}^{x^+}dx\,J_0(2\pi |\Delta|\sqrt{\sum m_i^2L_i^2}) \bigg] \nonumber\\
    \label{e:opticalLHS1}
    & = \frac{\lambda^2}{16\pi}\tilde Q\theta(\tilde Q^2)\sum_{\vec{\tilde m}\in\Lambda^n}\sinc(\pi \sqrt{s}\tilde Q|\vec{\tilde m}|).
\end{align}
In the first line, the 1 in the square brackets is the contribution from $\int dx\,\im\ln\mu^2/\Delta^2$. In the second line we exploited $\int_0^a dx \, J_0(\sqrt{a^2-x^2})=\sin(a)$, and $\sinc(x) \equiv \sin(x)/x$.  We also exchanged the sum over integers $\vec m$ (which are weighted by $L_i^2$ in the summand) with a sum over the lattice $\Lambda^n$ defined by the $n$ lengths $L_i^2$.  We make this last change to a sum over a lattice in anticipation of exploiting the Poisson summation formula over lattices \cite{JohnsonMcDaniel2012ADC}.  The Poisson summation over lattices is given by
\begin{align}
    \label{e:ps}
    \sum_{\vec{\tilde m}\in\Lambda^n}f(\vec{\tilde m}) = \frac{1}{\sqrt{\det \Lambda}}\sum_{\vec{\tilde k}\in\Lambda^{*n}}\tilde F(\vec{\tilde k}),
\end{align}
where $\Lambda^*$ is the lattice dual to $\Lambda$ and $\tilde F$ is the usual Fourier transform of $f$, $\tilde F(\vec {\tilde k} )\equiv \int d^n m \, e^{2\pi\ess i \ess \vec k\cdot \vec m} f(\vec {\tilde m})$.  Following exactly the method of performing the $n$ dimensional Fourier transform as shown in \appref{s:appsum} with now $f(\vec{\tilde m}) = \sinc(\pi \sqrt{s}\tilde Q|\vec{ \tilde m}|)$, we have that $\tilde F(\vec {\tilde k}) = \Omega_{2-n}\big( \frac s4\tilde Q^2-\tilde k^2 \big)^{\frac{1-n}{2}}\theta\big( \frac s4\tilde Q^2-\tilde k^2 \big)$.  Thus
\begin{multline}
    \label{e:opticalLHS}
    2\, \im \mathcal M = \frac{\lambda^2}{2(4\pi)^2\sqrt s}\theta(\tilde Q^2) \Omega_{2-n} \\ \times\frac{1}{\prod L_i}\sum_{\vec{\tilde k}\in\Lambda^{*n}}\big( \frac s4\tilde Q^2-\tilde k^2 \big)^{\frac{1-n}{2}}\theta\big( \frac s4\tilde Q^2-\tilde k^2 \big).
\end{multline}

Consider now the total cross section,
\begin{align}
    \sigma_{tot}
    & = \frac12\prod_{j=1}^2\sum_{\vec k_j\in\mathbb Z^n}\frac{1}{(2\pi)^n\prod L_i}\int\frac{d^mp_j}{(2\pi)^m2E_j} \nonumber\\
    & \qquad \times\lambda^2(2\pi)^4\prod L_i\delta(p_A^0+p_B^0-p_1^0-p_2^0) \nonumber\\
    & \qquad \times \delta^{(m)}(\vec p_A + \vec p_B - \vec p_1 - \vec p_2)\delta^{(n)}_{\vec k_A+\vec k_B,\vec k_1+\vec k_2}.
\end{align}
We may immediately collapse the $p_2$ integrals with the Dirac delta functions and the $k_2$ sums with the Kronecker deltas.  Then 
\begin{align}
    \sigma_{tot}
    & = \frac{\lambda^2}{2(2\pi)^2}\sum_{\vec k_1\in\mathbb Z^n}\frac{1}{\prod L_i}\int \frac{d^mp_1}{(2E_1)^2} \nonumber\\ 
    & \qquad\qquad\qquad\qquad\times \delta(\sqrt{s} - 2\sqrt{p_1^2 + \sum \frac{k_i^2}{L_i^2}+m^2}) \nonumber\\
    \label{e:opticalRHS}
    & = \frac{\lambda^2}{2(4\pi)^2\sqrt s}\theta(\tilde Q^2)\Omega_{2-n} \\
    & \qquad \times \frac{1}{\prod L_i}\sum_{\vec {\tilde k}\in\Lambda^{*n}}\big( \frac s4\tilde Q^2 - \vec{\tilde k}^2 \big)^{\frac{1-n}{2}} 
    \theta\big( \frac s4\tilde Q^2 - \vec{\tilde k}^2 \big). \nonumber
\end{align}
To arrive at the last line we integrate out the Dirac delta function, use $n+m=3$ to set $m = 3-n$, and change the sum over the integers to a sum over the lattice dual to the lattice from \eq{e:opticalLHS1}.

One can readily see that \eq{e:opticalLHS} and \eq{e:opticalRHS} are equal, and therefore the Optical Theorem (i.e.\ unitarity) is satisfied for our newly derived result for $n = 0,\,1,$ and $2$.  The $n=3$ case is more subtle, but for physical values of momenta, both $\im\mathcal M$ and $\sigma_{tot}$ diverge similarly.  

\section{Conclusions and Outlook}
In this work we computed the finite size correction to $2\rightarrow2$ scattering at NLO in a $\mathbb R^{1,m}\times T^n$ universe.  To do so, we introduced denominator regularization, derived an analytic continuation of the generalized Epstein zeta function, and gave the explicit result for $m+n=3$.  The denominator regularization naturally isolated the $1/\epsilon$ UV divergence and allowed for asymmetric finite sized spaces.  We performed a non-trivial check by confirming that our explicit result respects the Optical Theorem.  In performing the check, the generalized Poisson summation formula \cite{JohnsonMcDaniel2012ADC} allowed us to equate \eq{e:opticalLHS1} with \eq{e:opticalLHS}.  The equality of these two formulae is equivalent to a generalization of the conjecture by Hardy \cite{Hardy1978RamanujanTL}
\begin{align}
    \sum_{n=0}^{\lfloor x\rfloor} \frac{r_2(n)}{\sqrt{x-n}} = 
    2\pi\sqrt x \sum_{n=0}^\infinity r_2(n)\sinc(2\pi\sqrt{nx}),
\end{align}
where $r_2$ is the square counting function (in 2D) and $x$ is a positive non-integer, to
\begin{align}
    \frac{\pi^{\frac{1-m}{2}}}{2\sqrt x\Gamma\big(\frac{3-m}{2}\big)}\sum_{n=0}^{\lfloor x\rfloor} \frac{r_m(n)}{\sqrt{x-n}^{m-1}} = \sum_{n=0}^\infinity r_m(n)\sinc(2\pi \sqrt{nx}),
\end{align}
where $r_m$ is the square counting function in $m$ dimensions.

Although not shown here, we have explicitly checked that all momentum dependent subdivergences in the two loop 4 point function in $\phi^4$ theory cancel in denominator regularization; i.e.\ all divergences up to two loops in the 4 point function can self-consistently be absorbed in momentum-independent counter terms.  While we have not yet checked explicitly, very interesting future work includes showing that denominator regularization preserves Lorentz and gauge invariance and allows renormalization to all orders.  One should further be able to readily apply denominator regularization to problems in thermal field theory by replacing the integral over $\ell^0$ in \eq{eq:one} with a sum over Matsubara modes in the imaginary time formalism, and then capturing divergences using the analytic continuation of the Epstein zeta function \eq{e:epsteinanalytic}.  

Crucially we see that the $\mu$ dependence only resides in the $\mathbb R^{1,3}$ contribution.  Thus the running coupling from Callan-Symanzik is insensitive to finite size effects.  This is perhaps not surprising as Callan-Symanzik only captures the leading logarithmic behavior of the effective coupling for large scales, where the effective coupling comes from the resummation of bubble diagrams.  To determine the finite size effects on the coupling, then, one must perform the full resummation, $\lambda_{eff} = \lambda / \big( 1-\lambda\big(\bar V(s) + \bar V(t) + \bar V(u) \big)$.  We will fully explore the highly non-trivial qualitative and quantitative behavior for $\lambda_{eff}$ in future work \cite{Horowitz:2022}.

We believe the techniques developed here may be used in a number of other physical applications.  These techniques should easily allow for a computation of the finite size effective coupling in thermal field theory.  Also, there are a number of physical systems that display conformal symmetry, even non-relativistically, for which finite size system effects may play an interesting role \cite{Camblong:2003mz}.  Moreover, one ought to be able to compute the finite size corrections to critical exponents in the universality class of $\phi^4$ theory through the resummed 2 point function \cite{Cardy:1996xt}.  The latter may provide valuable insight, e.g., in detecting the critical endpoint of the QCD phase diagram from measurements of particle fluctuations in hadronic collisions \cite{Stephanov:1998dy,Luo:2017faz}.  

\section*{Acknowledgments}
WAH wishes to thank the South African National Research Foundation and the SA-CERN Collaboration for support and New Mexico State University for its hospitality, where part of the work was completed.  The authors wish to thank Matt Sievert, Alexander Rothkopf, Bowen Xiao, Stan Brodsky, Andrea Shindler, and Kevin Bassler for valuable discussions.

\appendix
\section{Analytic Continuation of the Generalized Epstein Zeta Function}
\label{s:appsum}
We would like to analytically continue the generalized We follow the starting steps of \cite{Elizalde:1997jv}.  One begins from the Poisson summation formula, \eq{e:ps}, applied to the generalized Eptein zeta function, \eq{e:epsteinzeta}.  We need to evaluate the Fourier transform of the generalized Epstein zeta function.  Let's consider the case in which we subtract a small imaginary part from $c$ such that we avoid the possibility of integrating through any poles.  Then for $\varepsilon>0$
\begin{multline}
    \int d^px\ess e^{2\pi\ess i \vec k\cdot \vec x}\big(a_i^2x_i^2+b_ix_i+c-i\varepsilon\big)^{-s} = \\
    e^{-2\pi \ess i \sum\limits_{i=1}^p\frac{k_i\ess b_i}{2a_i^2}}\frac{1}{\prod\limits_{i=1}^pa_i}\int d^p x' e^{2\pi \ess i \ess \vec k \cdot \vec x'}\big( \vec x'^2 + c' -i\varepsilon\big)^{-s},
\end{multline}
where $x'_i \equiv x_i + \frac{b_i}{2a_i^2}$ and $c' \equiv c - \sum_{i=1}^p \frac{b_i^2}{4a_i^2}$.  The remaining integral may be split into radial and angular parts,
\begin{align}
    \int & d^p x' e^{2\pi \ess i \ess \vec k \cdot \vec x'}\big( \vec x'^2 + c' -i\varepsilon\big)^{-s} \nonumber\\
    & = \int_0^\infty \rho^{p-1}d\rho(\rho^2+c'-i\varepsilon)^{-s}\int d\Omega_{p-1}e^{2\pi\ess i \ess k\ess \rho \cos\theta},
\end{align}
where $\rho\equiv|\vec x'|$, $k\equiv|\vec k|$, and $\Omega_p = 2\pi^{\frac{p+1}{2}}/\Gamma\big(\frac{p+1}{2}\big)$ is the solid angle of a $p$-dimensional sphere; $\Omega_2 = 4\pi$.  The angular integration evaluates for $k\ess\rho>0$ (which is always satisfied in our case) and $p>1$ to
\begin{multline}
    \int d\Omega_{p-2}\int_0^\pi\sin^{p-2}\theta d\theta e^{2\pi\ess i \ess k \ess \rho \cos\theta} \\
     = \frac{2\pi^{p/2}}{\Gamma(p/2)}{}_0F_1\big(;\frac{p}{2};-(\pi\ess k \ess \rho)^2\big),
\end{multline}
where ${}_0F_1(;a;z)$ is a usual generalized hypergeometric function.  One may check that for $p=2$ the above correctly reproduces $2\pi J_0(2\pi\ess k \ess \rho)$, where $J_\nu(z)$ is the usual Bessel function of the first kind.  One may then complete the evaluation through the use of
\begin{multline}
    \label{e:epsteinfourier}
    \int_0^\infty \rho^{p-1}d\rho(\rho^2+c'-i\varepsilon)^{-s}\frac{2\pi^{p/2}}{\Gamma(p/2)}{}_0F_1\big(;\frac{p}{2};-(\pi\ess k \ess \rho)^2\big) \\
    =\frac{2\pi^2}{\Gamma(s)}\Big( \frac{c'-i\varepsilon}{\vec k^2} \Big)^{\frac p4-\frac s2}K_{s-\frac p2}(2\pi |\vec k|\sqrt{c'-i\varepsilon})
\end{multline}
for $\Re s > p/2$, $d>1$, $c'\in\mathbb R$, $\varepsilon>0$, and $|\vec k|>0$, where $K_\nu(z)$ is the usual modified Bessel function of the second kind.

\eq{e:epsteinfourier} doesn't have an obvious $\vec k = \vec 0$ limit.  We must therefore separately evaluate the $\vec k = \vec 0$ Fourier mode.  One finds that for $\Re s > d/2 > 0$, $c'\in\mathbb R$, and $\varepsilon > 0$
\begin{align}
    \int d^p x' (\vec x'^2 + c' - i\varepsilon)^{-s} 
    = \pi^{p/2}\frac{\Gamma\big(s-\frac p2\big)}{\Gamma (s)}(c'-i\varepsilon)^{\frac p2-s}. \nonumber
\end{align}

Putting the pieces together we arrive at our master formula for the analytic continuation of the generalized Epstein zeta function:
\begin{widetext}
    \begin{multline}
        \label{e:epsteinanalytic}
        \sum_{\vec n\in\mathbb Z^p} (a_i^2 n_i^2 + b_i n_i + c-i\varepsilon)^{-s} = \frac{1}{a_1\cdots a_p}\frac{1}{\Gamma(s)} \bigg[ 
        \pi^{p/2}\Gamma\big( s - \frac p2\big) \Big( c - \sum \frac{b_i^2}{4a_i^2} - i\varepsilon \Big)^{\frac p2 - s} \\
        +2\pi^s\sideset{}{'}\sum_{\vec m\in\mathbb Z^p} e^{-2\pi\ess i \sum \frac{m_i b_i}{2a_i^2}}\bigg( \frac{c - \sum\frac{b_i^2}{4a_i^2} - i\varepsilon}{\sum \frac{m_i^2}{a_i^2}} \bigg)^{\frac p4-\frac s2} 
        K_{s-\frac p2}\bigg(2\pi \sqrt{(c-\sum\frac{b_i^2}{4a_i^2} - i\varepsilon)\big(\sum \frac{m_i^2}{a_i^2}\big)} \bigg)
        \bigg],
    \end{multline}
\end{widetext}
where $\sideset{}{'}\sum\limits_{\vec m\in\mathbb Z^p}$ indicates a sum over all integers $\vec m\in\mathbb Z^p$ \emph{except} for $\vec m = \vec 0$ and where the suppressed limits on the sums run from $i=1\ldots p$.  Notice that the contribution from $\vec m = \vec 0$ isolates the pole as we analytically continue $s\rightarrow p/2$.

One may numerically evaluate the $\sideset{}{'}\sum\limits_{\vec m\in\mathbb Z^p}$ in \eq{e:epsteinanalytic} more efficiently by combining the phases into cosines.  The speedup comes from evaluating a pure real expression and from drastically reducing the total number of summed terms.  The result is a sum over all the subsets of the set of numbers $\{1,\ldots,p\}$, known as the power set, $2^{[p]}$:
\begin{widetext}
    \begin{multline}
        \label{e:moreefficient}
        \sideset{}{'}\sum_{\vec m\in\mathbb Z^p} e^{-2\pi\ess i \sum \frac{m_i b_i}{2a_i^2}}\bigg( \frac{c - \sum\frac{b_i^2}{4a_i^2} - i\varepsilon}{\sum \frac{m_i^2}{a_i^2}} \bigg)^{\frac p4-\frac s2} 
        K_{s-\frac p2}\bigg(2\pi \sqrt{(c-\sum\frac{b_i^2}{4a_i^2} - i\varepsilon)\big(\sum \frac{m_i^2}{a_i^2}\big)} \bigg) \\
        = \sum_{s\in2^{[p]}}2^{|s|+1}\sum_{\substack{m_i=1 \\ i\in s}}^\infinity\prod_{i\in s}\Big( \cos(2\pi\ess x \ess m_i L_i p^i) \Big)\bigg( \frac{c - \sum\frac{b_i^2}{4a_i^2} - i\varepsilon}{\sum \frac{m_i^2}{a_i^2}} \bigg)^{\frac p4-\frac s2} K_{s-\frac p2}\bigg(2\pi \sqrt{(c-\sum\frac{b_i^2}{4a_i^2} - i\varepsilon)\big(\sum \frac{m_i^2}{a_i^2}\big)} \bigg)
        ,
    \end{multline}
\end{widetext}
where $|s|$ is the length of the current set of indices being summed over and the sums with the suppressed limits are over $i\in s$.

\bibliography{FSrefs}
\bibliographystyle{iopart-num}
\end{document}